\documentclass[twocolumn,aps,superscriptaddress,floatfix,final,showpacs,prb,eqsecnum]{revtex4}
\usepackage{amsmath}
\usepackage{graphicx}
\usepackage{amssymb}

\makeatletter
\usepackage{epsfig}
\usepackage{dcolumn}
\usepackage{bm}
\usepackage{amsfonts}
\usepackage{graphicx}

\usepackage{color}

\begin{document}

\title{Wigner-Mott scaling of transport near \\
the two-dimensional metal-insulator transition}

\author{M. M. Radonji\'{c}}
\affiliation{Scientific Computing Laboratory, Institute of Physics
Belgrade, University of Belgrade, Pregrevica 118, 11080 Belgrade,
Serbia}

\author{D. Tanaskovi\'{c}}
\affiliation{Scientific Computing Laboratory, Institute of Physics
Belgrade, University of Belgrade, Pregrevica 118, 11080 Belgrade,
Serbia}

\author{V. Dobrosavljevi\'{c}}
\affiliation{Department of Physics and National High Magnetic Field Laboratory,
Florida State University, Tallahassee, Florida 32306, USA}

\author{K. Haule}
\affiliation{Department of Physics and Astronomy, Rutgers University, Piscataway,
New Jersey 08854, USA}

\author{G. Kotliar}
\affiliation{Department of Physics and Astronomy, Rutgers University, Piscataway,
New Jersey 08854, USA}

\begin{abstract}
Electron-electron scattering usually dominates the transport in 
strongly correlated materials. It typically leads
to pronounced resistivity maxima in the incoherent regime around the
coherence temperature $T^{*}$, reflecting the tendency of carriers
to undergo Mott localization following the demise of the Fermi liquid.
This behavior is best pronounced in the vicinity of interaction-driven
(Mott-like) metal-insulator transitions, where the $T^{*}$ decreases,
while the resistivity maximum $\rho_{max}$ increases. Here we show
that, in this regime, the entire family of resistivity curves displays
a characteristic scaling behavior $\rho(T)/\rho_{max}\approx F(T/T_{max}),$
while the $\rho_{max}$ and $T_{max}\sim T^{*}$ assume a powerlaw
dependence on the quasi-particle effective mass $m^{*}$. Remarkably,
precisely such trends are found from an appropriate scaling analysis
of experimental data obtained from diluted two-dimensional electron
gases in zero magnetic fields. Our analysis provides strong evidence
that inelastic electron-electron scattering -- and not disorder effects
-- dominates finite temperature transport in these systems, validating
the Wigner-Mott picture of the two-dimensional metal-insulator transition.
\end{abstract}


\maketitle

\section{Introduction}

The physical nature of scattering processes which control transport
represents one of the most fundamental properties for any material.
At the lowest temperatures the thermal excitations are few, and elastic
impurity scattering dominates. Raising the temperature introduces
two basic pathways to modify transport. First, elastic scattering
can acquire a temperature dependence either through the modified screening
of the impurity potential, or through dephasing processes.\cite{leeramakrishnan,zalaPRB2001} This general
mechanism encapsulates the physical content of all ``quantum corrections''
-- both in the diffusive and the ballistic regime -- predicted within
the Fermi liquid framework. Indeed, careful and precise experiments
have confirmed the validity of this physical picture for many good
metals with weak disorder. \cite{leeramakrishnan} Physically, it relies on the existence
of long-lived quasiparticles within a degenerate electron gas.

The second route comes into play in instances where correlation effects
due to electron-electron interactions are significant. Here, the Fermi
liquid regime featuring degenerate quasiparticles is often restriced
to a very limited temperature range $T \ll T^{*}\ll T_{F}$, well below
the ``coherence temperature'' $T^{*}$, which itself is much smaller
than the Fermi temperature $T_{F}$. In such materials, which include
rare-earth intermetallics,\cite{thompsonPRB1985,mcelfreshPRB1990}
many transition metal oxides,\cite{basovprb2006} and several
classes of organic Mott systems,\cite{limelettePRL2003,kurosakiPRL2005,merinoPRL2008}
a broad intermediate temperature
regime emerges $T\sim T^{*}\ll T_{F}$ where \emph{inelastic} electron-electron
scattering dominates all transport properties. Such scattering directly
reflects the thermal destruction of Landau quasiparticles -- a
situation describing the demise of a coherent Fermi liquid.
In these materials, in the relevant temperature range, the electron-phonon scattering is much weaker than
the electron-electron one.

When a material is tuned to the vicinity of any metal-insulator
transition,\cite{mott-book90} both disorder and electron-electron
interactions are of \emph{a priori }importance. But which of these
two scattering mechanisms -- elastic or inelastic -- dominates the
experimentally relevant temperature range? Answering this question
should provide important clues as to which of the localization
mechanisms dominate in any given material. Unfortunately,
experimental systems permitting sufficiently precise tuning of
control parameters are generally rather few. An attractive class
of systems where a dramatic metal to insulator crossover is
observed in a narrow parameter range is provided by two
dimensional electron gases (2DEG), such as silicon MOSFETs or
GaAs/AlGaAs
heterostructures.\cite{abrahamsRMP2001,kravchenkoRPP2004,RMP20spivak10} 
One of the most striking features observed in these systems
is the pronounced resistivity drop on the metallic side of
the transition. While conventional, relatively weak temperature dependence
is found at high densities ($n\gg n_{c})$, very strong temperature
dependence is found near the critical density $n_c$, roughly in the same
density range $n_{c}\lesssim n\lesssim2n_{c}$ where other strong
correlation phenomena were observed, e.g. large $m^{*}$ enhancement.\cite{shashkinPRB2002}
Here, pronounced resistivity maxima are observed at $T\sim T_{max}(n)$, 
followed by a dramatic resistivity drop at lower temperatures, whose physical
origin remains a subject of much controversy and
debate.\cite{abrahamsRMP2001,kravchenkoRPP2004,RMP20spivak10}

In this paper we argue that the electron-electron scattering
dominates the transport in a broad concentration and temperature
range on the metallic side of the metal-insulator transition (MIT) in Si MOSFETS and
GaAs/AlGaAs heterostructures. This conclusion is reached by: 
(i) A detailed scaling analysis of the metallic resistivity curves; (ii) Establishing a similarity
in the transport properties of the 2DEG and well-studied strongly
correlated materials near the interaction-driven MIT; (iii) Making
a comparison of the resistivity curves in 2DEG with those in a
simple model of the Mott MIT. Our conclusions favor the
interaction-driven (Wigner-Mott) scenario\cite{neilson99prb,spivakPRB2001,pankovPRB2008,camjayiNATPHY2008,amaricci2010prb} of the MIT in 2DEG and
provide a guidance for the development of a microscopical theory of
incoherent transport in diluted 2DEG.

The remaining part of the paper is organized as follows. Our
phenomenological scaling of the experimental data is shown in
Sec.~II, and the analogy with strongly correlated three
dimensional (3D) materials is highlighted in Sec.~III. The scaling
analysis in a simple model of the Mott MIT is presented in Sec.~IV,
and Sec.~V contains the conclusion and discussion.


\section{Scaling analysis of the resistivity maxima}

The experimental data reveal well defined trends in the density dependence
of the resistivity maxima, suggesting a scaling analysis. While many
different scenarios for the metal-insulator transition predict some
form of scaling, its precise features may provide clues to what
mechanism dominates the transport.

\begin{figure}[b]
\begin{center}
\includegraphics[  width=6.6cm, keepaspectratio]{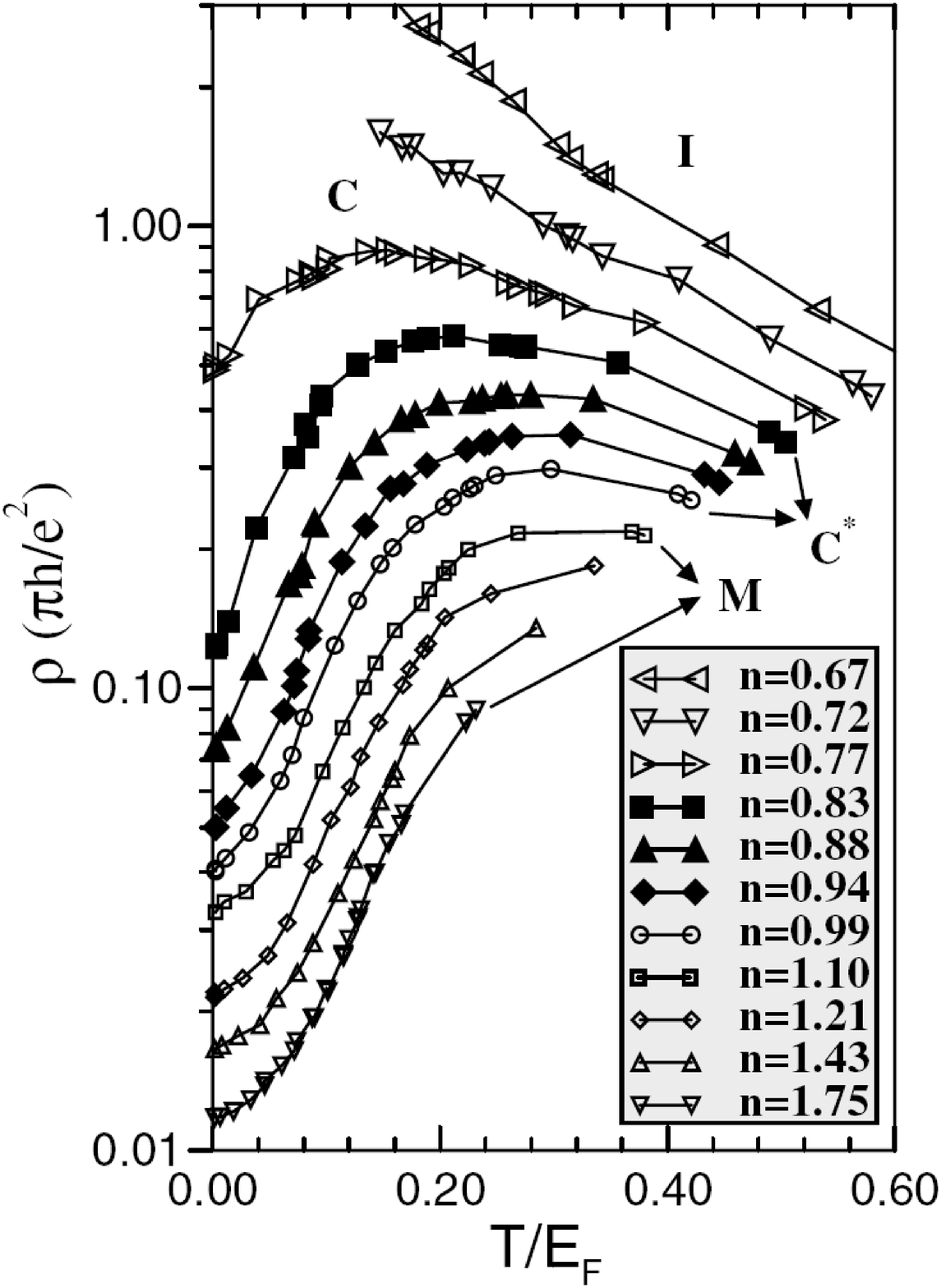}
\end{center}
\vspace{-0.5cm}
\caption{Resistivity as a function of temperature from the
experiments on Si MOSFET by Pudalov {\it et
al.}~(Ref.~\onlinecite{pudalovPHYSICA1998}). }
\label{pudalovdata}
\end{figure}

All of the curves displaying a resistivity maximum have an almost
identical shape [Fig.~\ref{pudalovdata}], strongly suggesting that unique physical
processes are responsible for a strong temperature dependence of
the resistivity in a large range of concentrations. The
resistivity maxima are typically observed at temperatures
comparable to the Fermi temperature, where a physical picture of
long-lived quasiparticles is no more valid.
Complementary experiments\cite{shashkinPRB2002,kravchenkoRPP2004} on the same
material have revealed that large effective mass $m^{*}$
enhancements are observed in the same density range. This behavior
is a clear signature of strong correlation effects which, in all
known examples, produce very strong inelastic electron-electron
scattering in the appropriate temperature range.
The electron-phonon scattering is negligibly small for
$T < T_F\lesssim 10$ K.\cite{andoRMP1982}
Since a strongly correlated system is typically characterized
by a single characteristic energy scale $T^* \sim (m/m^*) \,
T_F$, we expect the scaling function $f(x)$ to assume a universal
form, while the scaling parameters $T_{max}\equiv T^*$ and
$\rho_{max}$ to assume a simple, power-law dependence on the
effective mass $m^{*}.$
Guided by these observations, in this Section we introduce a
scaling ansatz and perform a scaling analysis of the resistivity
curves in Si MOSFETs and GaAs heterostructures.

\subsection{Phenomenological scaling hypothesis}

\begin{figure}[t]
\begin{center}
\includegraphics[  width=7.cm, keepaspectratio]{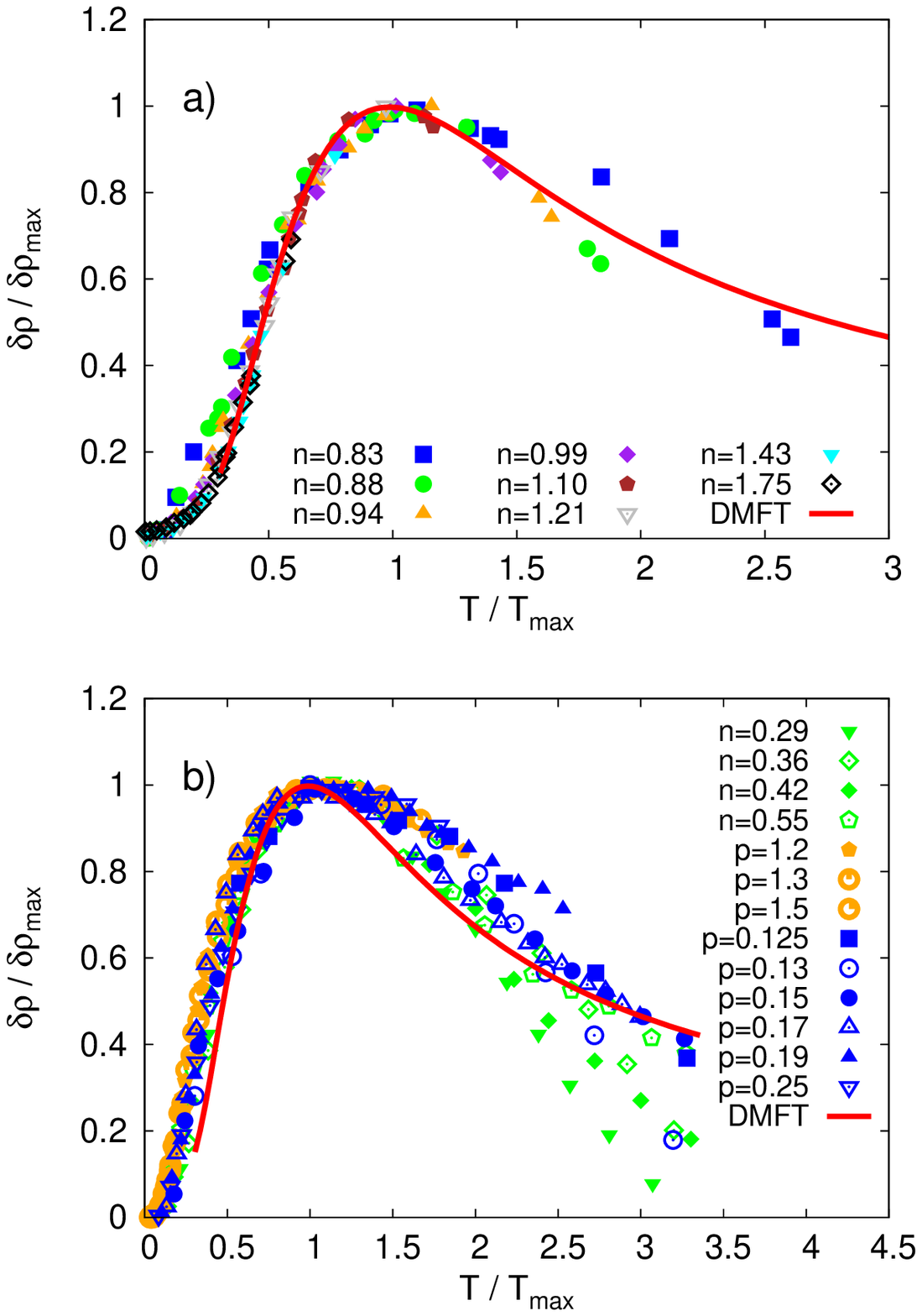}
\end{center}
\vspace{-0.5cm}
\caption{(Color online) Scaled resistivity as a function of scaled temperature
for different electron (hole) concentrations, for Si MOSFET (a)
and GaAs heterostructures (b). The experimental data are taken
from Ref.~\onlinecite{punnoosePRL2001} (MOSFETs),
Ref.~\onlinecite{haneinPRL1998} (p-GaAs/AlGaAs, blue symbols),
Ref.~\onlinecite{lillyPRL2003} (n-GaAs/AlGaAs, green symbols), and
Ref.~\onlinecite{gaoPRL2005} (p-GaAs, orange symbols). The
solid line is the scaling function obtained for a simple model of
the MIT (see Sec.~IV).}
\label{scaledexp}
\end{figure}

In accordance to what is typically found in other examples of strongly
correlated metals with weak to moderate disorder,\cite{limelettePRL2003} we expect the
resistivity to assume an additive form,
 $\rho(T)=\rho_{o}+\delta\rho(T)$.
Here, $\rho_{o}$ is the residual resistivity due to impurity scattering,
and the temperature-dependent contribution $\delta\rho(T)$
is expected to be dominated by inelastic
electron-electron scattering.
Based on these general considerations,
we propose that the temperature-dependent term assumes a scaling form
\begin{equation}\label{scaling_ansatz}
\delta\rho(T)=\delta\rho_{max}f(T/T_{max}),
\end{equation}
where $\delta\rho_{max}=\rho_{max}-\rho_{o}$.

\begin{figure}[h]
\begin{center}
\includegraphics[  width=7.cm, keepaspectratio]{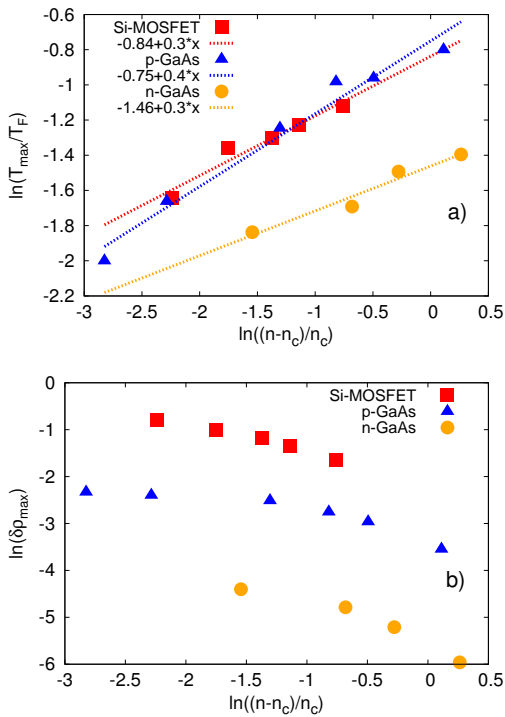}
\end{center}
\vspace{-0.5cm}
\caption{(Color online) $T_{max}$ normalized to Fermi temperature (a),
and maximal resistivity $\delta \rho_{max}= \rho_{max}-\rho_o$ in
units $\pi h/e^2$ (b), as a function of a reduced density.
The data are taken from Refs.~\onlinecite{punnoosePRL2001,haneinPRL1998,lillyPRL2003}.
}
\label{n-nc_scaling}
\end{figure}

To test this phenomenological scaling hypothesis, we perform a
corresponding analysis of experimental data in several systems
displaying 2D-MIT.
We start with the Si MOSFET data\cite{pudalovPHYSICA1998} analyzed in Ref.~\onlinecite{punnoosePRL2001}.
We concentrate on metallic curves below the separatrix C. In the
range of concentrations, $0.83 < n < 1.10$, the resistivity curves
have a clear maximum, and nicely collapse with the proposed scaling ansatz,
Fig.~\ref{scaledexp}(a). In fact, we can use the scaling ansatz
to collapse also the data for $1.21 < n < 1.75$,
where $T_{max}$ and $\rho_{max}$ are determined from the least square fit
to the scaling curve.
Clearly all eight resistivity curves belong to
the same family (have the same functional form), and thus must be
explained by a \emph{single dominant transport mechanism}.
This conclusion is even more convincing if we apply the same analysis
to several different materials, including an ultra high mobility GaAs
sample, Fig.~\ref{scaledexp}(b). While the diffusive physics cannot possible
apply in such a broad parameter range, we see that the scaling
form we propose proves to be an extremely robust feature of all
available 2D-MIT systems. This result is very significant, because
disorder effects must be significantly weaker in these ultra-clean
materials, while the interaction effects are expected to be even
stronger.

\subsection{Critical behavior of the Wigner-Mott scaling}

\begin{figure}[h]
\begin{center}
\includegraphics[  width=6.cm, keepaspectratio]{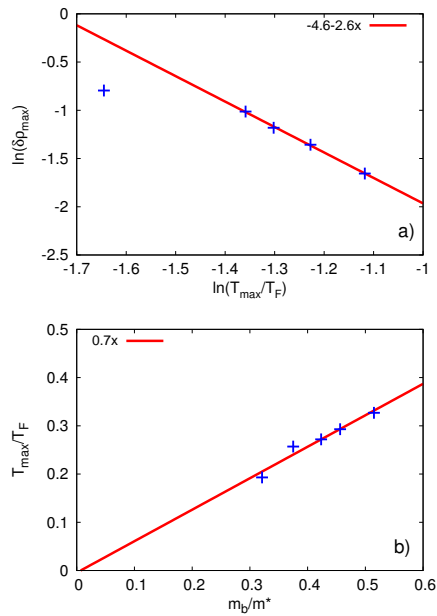}
\end{center}
\vspace{-0.5cm}
\caption{(Color online) (a) Maximum resistivity $\delta \rho_{max}=
\rho_{max}-\rho_o$ as a function of $T_{max}$.  (b) $T_{max}$ as a
function of inverse effective mass $m^*$. $m_b$ is the band mass in
Si MOSFETs. The data are taken from
Refs.~\onlinecite{punnoosePRL2001} and \onlinecite{prusPRB2003}. }
\label{Tmax_vs_m}
\end{figure}

Having demonstrated data collapse, we are now in a position to
examine the critical behavior of the relevant crossover scale. We
thus examine the behavior of $T_{max}$ and $\rho_{max}$ as a
function of reduced concentration $(n-n_c)/n_c$ and effective mass
$m^*$ (as determined by complementary experiments).

For different realizations of 2DEG, $T_{max}$ shows approximately
power law dependence on the reduced concentration [Fig.~\ref{n-nc_scaling}(a)] and
even the exponents are similar. $T_{max}$ in our physical picture
has a clear physical interpretation as a coherence temperature -
the temperature when the inelastic electron-electron scattering time
becomes comparable to $\hbar/E_F$, leading to incoherent
transport. The resistivity maximum, however, shows less universal
form. It varies a lot in different physical systems. This does not
come as a surprise since the resistivity shows nonuniversal
features also in three dimensional strongly correlated materials
near the Mott transition. We discuss in detail the analogy with
the Mott systems in Secs.~III and IV.

In a Si MOSFET the resistivity maximum $\delta \rho_{max}=
\rho_{max}-\rho_o$ shows power law dependence on $T_{max}$ in a
fairly broad concentration range [Fig.~\ref{Tmax_vs_m}(a)]. We further analyze
the critical behavior for Si MOSFET using the data for the
effective mass as determined by Shashkin {\it et
al.}\cite{shashkinPRB2002} from magnetoresistance measurements in
a parallel magnetic field. We find that $T_{max}$ is inversely
proportional to the effective mass $m^*$. This behavior is typical
to all systems near the Mott MIT, where the coherence temperature
is inversely proportional to the effective mass, as a landmark of
strong correlations.

\subsection{Breakdown of the diffusion mode scaling}

We have successfully collapsed resistivity curves in a broad
temperature and concentration range and for several physical
realizations of 2DEG. The physical picture behind the proposed
scaling is that the 2D MIT is an interaction-driven (Wigner-Mott)
MIT,\cite{neilson99prb,spivakPRB2001,pankovPRB2008,camjayiNATPHY2008,amaricci2010prb} and that the dominant temperature dependence in the
resistivity originates from strong electron-electron scattering.
\begin{figure}[b]
\begin{center}
\includegraphics[  width=8cm, keepaspectratio]{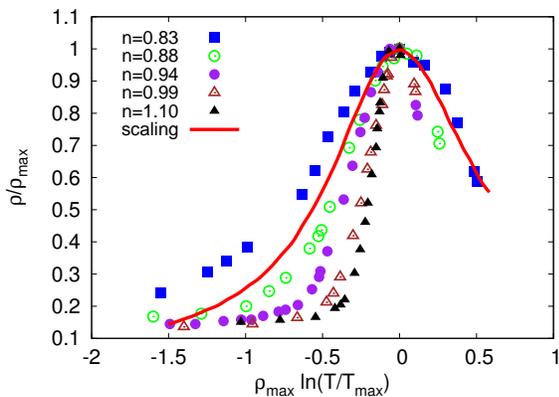}
\end{center}
\vspace{-0.5cm} \caption{(Color online) Resistivity as a function of temperature scaled as in
Ref.~\cite{punnoosePRL2001}. Red solid line is the calculated scaling curve.}
\label{PFscaling}
\end{figure}
Another proposed scenario envisions disorder as the principal
driving force for localization,\cite{punnoosePRL2001,hamiltonPRL2002} while the interactions are most
important above the critical density and at low temperatures,
where they suppress the tendency to localization. An appropriate
theory, based on a Fermi liquid framework,\cite{punnoosePRL2001} has predicted that a
resistivity maximum should be observed on the metallic side, with
the resistivity assuming the scaling form
\begin{equation}\label{PF_scaling}
\rho(T)/\rho_{max}=f[\rho_{max}\ln(T/T_{max})].
\end{equation}
Here $f(x)$ is a universal scaling function predicted by theory.
The authors point out, though, that this prediction is expected to
be valid only within the diffusive regime, where the thermal
energy $k_{B}T$ is smaller than the elastic scattering rate
$\hbar/\tau$. According to this picture, a different (ballistic)
mechanism for transport is expected outside the diffusive regime,
presumably leading to a different temperature dependence, so the
proposed scaling no longer holds. This analysis was applied to the
experimental data of Ref.\onlinecite{pudalovPHYSICA1998}, but was
accordingly restricted to only three densities closest to the
transition. Indeed, if the scaling formula is applied in a broader
range of concentrations, the resistivity curves clearly do not
collapse [Fig.~\ref{PFscaling}]. While the Fermi liquid renormalization group
calculations are very important in order to answer a fundamental
question of necessary conditions for a true MIT at zero
temperature, our analysis emphasizes that the understanding of
various diluted 2DEG in a broad range of parameters requires the
physics beyond the conventional Fermi liquid framework.

\section{Scaling in 3D materials}

\begin{figure}[b]
\begin{center}
\includegraphics[  width=7.cm, keepaspectratio]{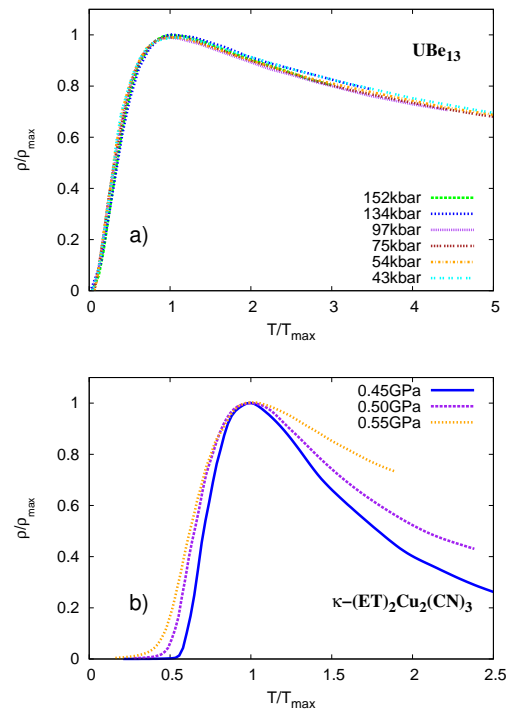}
\end{center}
\vspace{-0.5cm} \caption{(Color online) Scaled resistivity curves
for $\mbox{UBe}_{13}$ (a) and $\kappa$-$(\mbox{ET})_2 \mbox{Cu}_2
(\mbox{CN})_3$ (b), for different external pressure. The data are
taken from Refs.~\onlinecite{mcelfreshPRB1990} and
~\onlinecite{kurosakiPRL2005}.}
\label{3Dmaterials}
\end{figure}

The strong temperature dependence of resistivity is a well known feature of
many strongly correlated materials. A pronounced resistivity maximum is
observed in heavy fermions,\cite{thompsonPRB1985,mcelfreshPRB1990} and
charge-transfer organic salts,\cite{limelettePRL2003,kurosakiPRL2005,merinoPRL2008}
where the correlation strength is tuned by applying an external pressure.
The essential mechanism of transport in these materials relies on strong
inelastic electron-electron scattering, and the Fermi liquid
behavior is restricted to the lowest temperatures. As the temperature increases,
the electron mean free path becomes comparable to, or smaller than the lattice spacing,
and the transport becomes incoherent. The electron-phonon scattering is here much weaker
than the electron-electron one.
The temperature of the resistivity maximum can
be taken as a definition of the coherence temperature $T^*$. It is inversely proportional
to the effective mass, and much smaller than the bare Fermi temperature,
$T^* \sim (m_b/m^*) \, T_F$. The same scaling ansatz
as given by Eq.~(\ref{scaling_ansatz}) was used to collapse the resistivity curves
for $\mbox{CeCu}_6$ already in an early paper
by Thompson and Fisk.\cite{thompsonPRB1985}

Here we illustrate the similarity in transport properties of these
systems and 2DEG by scaling  the resistivity data for heavy
fermion $\mbox{UBe}_{13}$ from Ref.~\onlinecite{mcelfreshPRB1990}
[Fig.~\ref{3Dmaterials}(a)], and for a charge-transfer conductor
$\kappa$-$(\mbox{ET})_2 \mbox{Cu}_2 (\mbox{CN})_3$ [Fig.~\ref{3Dmaterials}(b)].
The collapse of the resistivity curves is excellent for
$\mbox{UBe}_{13}$, and well-defined trends are seen in $\kappa -
(\mbox{ET})_2 \mbox{Cu}_2 (\mbox{CN})_3$. 
Remarkable similarity in
resistivity curves in such diverse physical systems like Si MOSFETs,
GaAS heterostructures, heavy fermions and charge-transfer organic
conductors is, in our view, a manifestation of the same physical
processes in the vicinity of the interaction-driven MIT.

\section{Scaling in the microscopic model of the interaction-driven MIT}

\begin{figure}[t]
\begin{center}
\includegraphics[  width=7 cm, keepaspectratio]{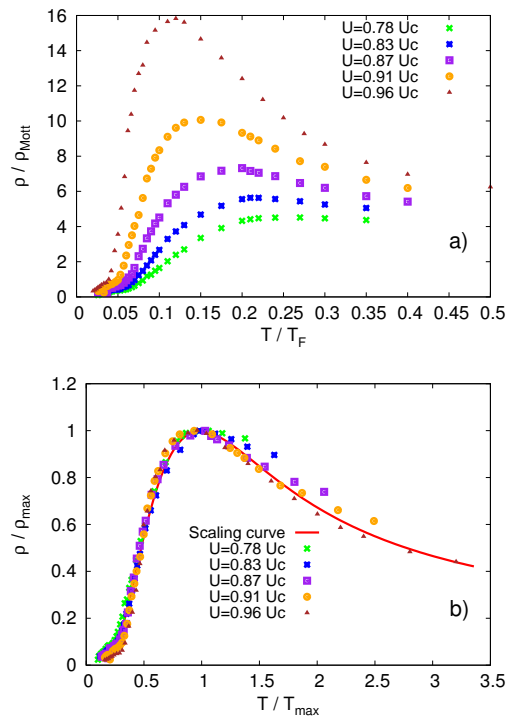}
\end{center}
\vspace{-0.5cm} \caption{(Color online) (a) Resistivity as a
function of temperature for several interaction strengths in the
half-filled Hubbard model solved within the DMFT. The resistivity
is normalized to the Mott limit value, which corresponds to the scattering
length of one lattice spacing. (b) Scaled resistivity curves.}
\label{DMFTscaling}
\end{figure}

Having phenomenologically established precise and well defined
scaling behavior of the experimental curves on the metallic side
of the 2D MIT for temperatures near $T^*$, we now address its
microscopic origin. More precisely, we would like to understand
just how robust this result is. Does it depend on subtle details
describing the interplay of disorder and interactions of 2DEG
materials, as suggested in Ref.~\onlinecite{waintalPRB2010}, or is
it a generic feature of strong correlation near interaction-driven
MIT. To answer this important question we deliberately focus on
the simplest microscopic model for interaction-driven MIT: The
clean single-band Hubbard model at half-filling. Accurate and
quantitatively precise results can be obtained for
temperature-dependent transport for this model within the DMFT
approximation.\cite{georgesRMP1996} While the DMFT reproduces
Fermi liquid behavior at the lowest temperatures, it is
particularly useful in the studies of "high temperature"
incoherent transport. Results of such calculation, obtained by the
Continuous Time Quantum Monte Carlo (CTQMC) impurity
solver\cite{wernerPRL2006,haulePRB2007} followed by the analytical
continuation by the Maximum Entropy Method\cite{maierRMP2005}, can
be analyzed using precisely the same scaling procedure we proposed
for experimental data. We concentrate on the metallic phase of the
Hubbard model with the interaction parameter $U$ smaller than the
value at the critical end-point $U_c$. The resistivity curves
[Fig.~\ref{DMFTscaling}(a)] have qualitatively the same form as in 2DEG. The
resistivity sharply increases with temperature, reaches a maximum
and than decreases. The temperature of resistivity maximum
decreases as the system approaches the MIT.

\begin{figure}[t]
\begin{center}
\includegraphics[  width=6.5 cm, keepaspectratio]{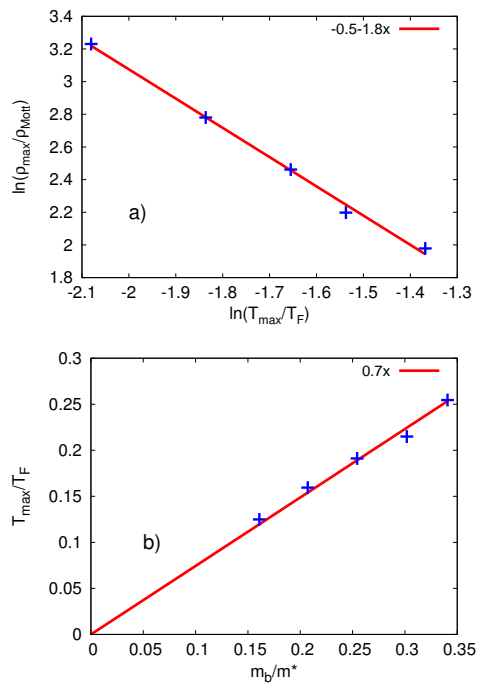}
\end{center}
\vspace{-0.5cm} \caption{(Color online) (a) Maximum resistivity as
a function of the corresponding temperature from the DMFT solution
of the Hubbard model. (b) $T_{max}$ as a function of the inverse
effective mass.}
\label{DMFT_Tmax_vs_m}
\end{figure}

Most remarkably, precisely the same scaling form as in 2DEG is found to
describe all resistivity curves close to the Mott
transition [Fig.~\ref{DMFTscaling}(b)]. In addition, we find that the scaling parameters
$T_{max}$ and $\rho_{max}$ again display a power law dependence on
the effective mass [Fig.~\ref{DMFT_Tmax_vs_m}], and even the exponents are similar.
Finally, we contrast the DMFT
scaling function with that obtained from 2DEG experiments. We find
surprisingly accurate agreement between the DMFT prediction for
the scaling function $f(x)$ and experimental data on all available
materials [Fig.~\ref{scaledexp}]. We emphasize, however, that our scaling
hypothesis is valid only in the metallic phase for $U<U_c$ and for
temperatures comparable to $T^* \sim 1/m^*$. It should be
contrasted with the scaling near the critical end-point
$(U_c,T_c)$,\cite{kotliarPRL2000,limeletteScience2003} or the
proposed quantum critical scaling in the high-temperature 
regime above the critical end-point.\cite{terletskaPRL2011}

We should point out that for this model, the proposed resistivity scaling is
not valid at the lowest temperatures $T \ll T_{max}$, deep within the
Fermi liquid region: According to the Kadowaki-Woods relation, here
$\rho \approx A T^2$ where $A \sim 1/m^{*2} \sim 1/T_{max}^2$, and the
scaling is violated if the resistivity is scaled by $\rho_{max}$.
For $T \gtrsim 0.3 T_{max}$ the collapse of the resistivity curves is excellent, [Fig.~\ref{DMFTscaling}(b)], and we define the DMFT scaling curve for this temperature range.
This is also the reason for the deviations in the scaling in
Fig.~\ref{3Dmaterials}(b) for $\kappa$-organics, the materials whose properties
are described remarkably well within the Hubbard
model.\cite{limelettePRL2003,merinoPRL2008} In the Anderson
lattice model, on the other hand, the resistivity maximum does not
change much near the MIT and it saturates approximately to the
value which corresponds to the scattering length of one lattice
spacing (Mott limit). In this case our scaling ansatz is valid in
the whole temperature range up to $T=0$,\cite{tanaskovicPRB2011} and
the collapse of the resistivity curves seen in the experiments is
excellent [Fig.~\ref{3Dmaterials}(a)].

Microscopic theory of the 2DEG should also include nonlocal correlations
which are neglected in a simple DMFT approach. 
A more realistic extended Hubbard model displays a two-stage Wigner-Mott localization.\cite{camjayiNATPHY2008,amaricci2010prb}  The metal-insulator transition in this model is found in the region with already developed nonlocal charge correlations. In the immediate critical regime, the critical behavior can be represented by an effective Hubbard model, partially justifying the success of the present modeling. 
The existence of a coherence scale $T^*$ which vanishes at the onset of
charge order is also found in the 2D extended Hubbard  model solved by finite-$T$ Lanczos
diagonalization.\cite{CanoCortesPRL2010} This result is relevant for quarter-filled layered organic
materials, which further supports the importance and generality of the ideas presented here.

\section{Conclusion and discussion}

In this paper we argued that the emergence of resistivity maxima
upon thermal destruction of heavy Fermi liquids should be regarded
as a generic phenomenon in strongly correlated systems. We
demonstrated that the resulting family of resistivity curves
typically obeys a simple phenomenology displaying scaling
behavior. Our detailed model calculations show that all of the
qualitative and even quantitative features of this scaling
phenomenology are obtained from a microscopic model of heavy
electrons close to the Mott metal-insulator transition.
We should stress, however, that the proposed scaling behavior
obtains - both in our theory and in experiments - only within the
metallic regime not too close to the transition and the
temperature regime around the resistivity maxima. In contrast,
earlier experiments focused on the immediate vicinity of the
metal-insulator transition, where different "quantum critical"
scaling was found.\cite{kravchenko95,abrahamsRMP2001,popovic97prl,gang2prl} Remarkably,
precisely such behavior was
also found in very recent studies of quantum critical transport
near interaction-driven transitions,\cite{terletskaPRL2011} but
this was identified in a different parameter regime than the one
studied in the present paper.

Our results provide compelling evidence that several puzzling
aspects of transport in low density two-dimensional electron gases
in zero magnetic fields can be understood and explained within the
Wigner-Mott scenario of strong correlation.\cite{neilson99prb,spivakPRB2001,pankovPRB2008,camjayiNATPHY2008,amaricci2010prb}
This physical picture views the strong
correlation effects in the low density 2DEG as the primary driving
force behind the transition, and additional disorder effects as
less significant, secondary processes. In the Wigner-Mott picture
the insulator essentially consists of interaction-localized magnetic moments. 
Remarkably, the magneto-capacitance measurements of Prus {\it et al.}\cite{prusPRB2003} 
show that the behavior characteristic of localized magnetic
moments, $\chi(T)/n\approx g\mu_{B}^{2}/T$, is seen near the
critical density, while only weak Pauli-like temperature dependence was observed at higher
density. Very recent experiments on Si MOSFETs find that the thermopower diverges 
near the MIT.\cite{mokashi2011} The authors argue that divergence
of the thermopower is not related to the degree of disorder and reflects
the divergence of the effective mass at a disorder-independent
density, behavior that is typical in the vicinity of an interaction-induced phase transition.
Additional hints supporting this physical picture of 2D MIT are provided by existing first principle
Quantum (diffusion) Monte Carlo results for the low density 2DEG of
Ceperley\cite{ceperley80prl} and others.\cite{tanatar95prl,waintalPRB2006,waintalPRB2010} These calculations find that the correlated metallic
state has an ``almost crystalline'' structure, thus having a very
strong short range charge-order (as seen, for example, in the
density correlation function).

Within the physical picture that we propose, the
inelastic electron-electron scattering takes central stage,\cite{aguiarEPL2004,andreevPRL2011}
in contrast to disorder-dominated scenarios,
where the interaction effects mainly introduce the temperature
dependence of {\em elastic} electron-impurity scattering.\cite{zalaPRB2001} The two
physical pictures describe two completely different scattering
processes, which are expected to be of relevance in complementary
but in essentially non-overlapping parameter regimes. Indeed,
inelastic scattering dominates only outside the coherent
Fermi-liquid regime, which in good metals happens only at fairly
high temperatures. In strongly correlated regimes that we consider, the
situation is different. Here the Fermi liquid coherence is found
only at very low temperatures $T <T^* \ll T_F$, behavior which is
generally observed in all system with appreciable effective mass
enhancement. The results presented in this paper
provide precise and detailed characterization of this incoherent regime,
revealing remarkable coincidence of trends observed in the
experiment to those found from the Wigner-Mott picture of the
interaction-driven metal-insulator transition. 
Our scaling ansatz is proposed based on the physical arguments and
the experimental data. While consistent with simple model calculations for strongly correlated 
electronic systems, our work does not directly address specific microscopic
mechanism responsible for current dissipation, a process that in 2DEG systems should be
facilitated by impurities and imperfections.\cite{andreevPRL2011} Still, it provides 
very strong motivation to develop a more realistic microscopic theory of incoherent transport 
in the strongly correlated regime of diluted 2DEG.
This important task remains a challenge for future work.

\begin{acknowledgments}The authors thank A. Punnoose and A. M. Finkel-
stein for usefull discussions.
D.T. and M.M.R. acknowledge support from the Serbian Ministry of Education and Science under
project No.~ON171017. V.D. was supported by the National High Magnetic
Field Laboratory and the NSF Grant No.~DMR-1005751, G.K. by the NSF Grant No.~DMR-0906943,
and K.H. by the NSF Grant No.~DMR-0746395.
Numerical simulations were run on the AEGIS e-Infrastructure, supported in
part by FP7 projects EGI-InSPIRE, PRACE-1IP and HP-SEE.
\end{acknowledgments}

\bibliographystyle{apsrev}

\end{document}